\documentclass[aps,prl,twocolumn,superscriptaddress]{revtex4-2}
\usepackage{amsmath}
\usepackage{amssymb}
\usepackage{amsthm}
\usepackage{graphicx}
\usepackage{todonotes}
\usepackage[hidelinks]{hyperref}
\usepackage[capitalise]{cleveref}

\newtheorem{theorem}{Theorem}

\DeclareMathOperator{\Tr}{Tr}
\newcommand{\indep}{\mathrel{\perp\!\!\!\perp}}

\begin{document}

\title{Existence of bound secrecy in six-state quantum key distribution}
\date{\today}

\author{Shin Sun}
\affiliation{Networked Quantum Devices Unit, Okinawa Institute of Science and Technology Graduate University, Okinawa, Japan}
\affiliation{LQUOM, Inc., Tokiwadai, Hodogaya-ku, Yokohama, Kanagawa, Japan}

\author{Daniel Bhatti}
\affiliation{Networked Quantum Devices Unit, Okinawa Institute of Science and Technology Graduate University, Okinawa, Japan}

\author{David Elkouss}
\affiliation{Networked Quantum Devices Unit, Okinawa Institute of Science and Technology Graduate University, Okinawa, Japan}

\begin{abstract}
 Quantum key distribution exploits quantum correlations to establish secrecy between distant parties. Although two-way classical communication can substantially improve its noise tolerance, whether arbitrarily sophisticated classical postprocessing can extract all the secrecy available in quantum correlations remains unresolved. Here we establish an exact limit on classical secret-key distillation in the six-state quantum key distribution protocol. We prove that no classical postprocessing can yield a positive asymptotic secret-key rate when the quantum bit-error rate reaches $q^*=(5-\sqrt{5})/10\approx27.64\%$, even when allowing arbitrary interactive communication, nonlinear operations and local private randomness. Together with known protocols achieving positive key rates below this threshold, our result establishes the exact noise tolerance of the six-state protocol under classical postprocessing and resolves a longstanding conjecture. Remarkably, the impossibility holds even in a regime where the underlying quantum state remains entangled, demonstrating that quantum entanglement does not necessarily translate into distillable secrecy through classical communication alone.
\end{abstract}

\maketitle
Quantum key distribution (QKD) protocols allow communicating parties to establish information-theoretically secure keys. For the secret key extraction of QKD protocols, different postprocessing techniques can be used to increase the key rate in certain noise regimes, with the noise commonly parametrized by $q$, the quantum bit error rate (QBER).

For the six-state protocol~\cite{brussOptimalEavesdroppingQuantum1998}, the standard one-way hashing rate
vanishes at $q\approx12.6\%$, while degenerate coding increases the tolerable QBER to approximately $q\approx12.7\%$~\cite{loProofUnconditionalSecurity2001}. The noise tolerance for one-way protocols can be improved with postprocessing \cite{krausLowerUpperBounds2005,rennerInformationtheoreticSecurityProof2005, renes2007noisy,kernImprovedOnewayRates2008}, with the noise tolerance increasing to $q\approx 14.59\%~$\cite{kernImprovedOnewayRates2008}. If the users instead employ two-way classical key distillation, i.e. advantage distillation, the noise tolerance increases further \cite{gottesmanProofSecurityQuantum2003,watanabe2007key,tupkary2022improved,duAdvantageDistillation2025}. With advantage distillation techniques positive-rate secret key distillation is possible below $q^* = (5-\sqrt{5})/10 \approx 27.64\%$~\cite{chauPracticalSchemeShare2002}. These results establish the achievability of the key distillation for both (classical) one-way and two-way postprocessing.
Allowing quantum processing
before measurement permits entanglement distillation, and hence
secret-key generation, for every
$q<1/3$~\cite{bennettPurificationNoisyEntanglement1996,loUnconditionalSecurityQuantum1999}. One round of an entanglement distillation protocol   \cite{deutsch1996quantum} is equivalent to advantage distillation \cite{murtaKeyRatesQuantum2020}. However, even a single round of entanglement distillation, followed by classical
advantage distillation, raises the noise tolerance above
$q^*$~\cite{sunEnhancingDistillabilitySecret2025}.
On the other hand, it is also known that secret key distillation is not possible for $q\geq1/3$ for the isotropic source considered below, by separability~\cite{horodeckiSeparabilityMixedStates1996,
curtyEntanglementPreconditionSecure2004}.

Nevertheless, it is not known whether secret key distillation with classical postprocessing is possible in the noise range $q^* \leq q < 1/3$. Numerical and analytical evidence indeed supports the conjecture that secret key distillation is impossible even with unbounded two-way classical communication. This conjecture is known as the \emph{bound secrecy conjecture}~\cite{baeKeyDistillationQuantum2007,khatriNumericalEvidenceBound2017}.
Bae and Ac\'in showed that standard advantage distillation followed by
one-way key extraction yields zero key for
$q\geq q^*$~\cite{baeKeyDistillationQuantum2007}.
Myhr et al.\ traced this limit to the symmetric extendibility of the state
after advantage distillation~\cite{myhrSymmetricExtensionTwoway2009}.
Khatri and L\"utkenhaus extended the analysis to more general code-based
postselection, obtaining analytical results for specific code families
and numerical evidence for others~\cite{khatriNumericalEvidenceBound2017}.
More recently, Tsurumaru established the impossibility of secret-key distillation beyond $q^*$ for a broad class of adaptive two-way linear postprocessing protocols~\cite{tsurumaruLimitsAdaptiveLinear2026}, providing further analytical support for the bound-secrecy conjecture.

In this work, we resolve the bound-secrecy conjecture for the six-state QKD protocol by constructing an explicit attack that renders secret-key distillation impossible for $q\geq q^*$. Our impossibility result holds for arbitrary interactive classical postprocessing, allowing unrestricted two-way communication, nonlinear operations, and local private randomness. Together with previous achievability results, our converse establishes the exact noise threshold for secret-key distillation in the six-state protocol, thereby revealing a fundamental limit of QKD that cannot be overcome by any classical postprocessing strategy.

\textit{Source and protocol model.---}
To prove the converse, it suffices to find one collective attack yielding zero key that is compatible with the observed six-state quantum bit-error rate $q$.
Let Alice and Bob's
qubit systems $A$ and $B$ share
\begin{align}
    \rho_q^{AB}
    &= (1-2q)\lvert\Phi^+\rangle\langle\Phi^+\rvert
       +\frac q2 I_4, \label{eq:quantum_source}\\
    \lvert\Phi^+\rangle
    &=\frac{\lvert00\rangle+\lvert11\rangle}{\sqrt2},
    \qquad 0\le q\le\frac12, \nonumber
\end{align}
where $I_4$ is the identity on the two-qubit space. Eve holds a purification
in a four-dimensional system $E$, with Hilbert space $\mathcal H_E$.
Alice and Bob measure in the computational basis before any classical
postprocessing. Their outcomes $X,Y\in\{0,1\}$ and Eve's remaining system
then have the classical--classical--quantum state
\begin{align}
    \omega_{XYE}
    &=\sum_{x,y\in\{0,1\}}p_{xy}\lvert xy\rangle\langle xy\rvert
      \otimes\lvert e_{xy}\rangle\langle e_{xy}\rvert,
      \label{eq:ccq_state}\\
    p_{00}=p_{11}&=\frac{1-q}{2},
    \qquad p_{01}=p_{10}=\frac q2. \nonumber
\end{align}
Here $p_{xy}=\Pr[X=x,Y=y]$, and $\lvert e_{xy}\rangle$ is Eve's normalized
conditional state whenever $p_{xy}>0$.

The protocol uses $n\ge1$ independent copies of \cref{eq:ccq_state}.
Write $X^n=(X_1,\ldots,X_n)$ and $Y^n=(Y_1,\ldots,Y_n)$ for the raw
strings, and $E^n=E_1\cdots E_n$ for Eve's $n$ quantum registers. Alice and
Bob may use arbitrary classical local processing and authenticated public
discussion~\cite{maurerSecretKeyAgreement1993}. Their private randomness is independent of the source and of
each other; any public randomness is also independent of the source and the
private randomness. No initially shared secret is supplied as a
key-generation resource.

Let $T$ be the complete public transcript, including public randomness,
stopping and acceptance decisions, and the output length. On the public
event $\mathrm{pass}$, assumed to have positive probability, the outputs
are $K_A,K_B\in\{0,1\}^{\ell}$ for a publicly specified fixed integer
$\ell\ge1$. Let $\sigma_{K_AE^nT\mid\mathrm{pass}}$ denote the normalized
accepted output state and $\sigma_{E^nT\mid\mathrm{pass}}$ its marginal.
We use correctness and trace-distance secrecy
errors~\cite{rennerInformationtheoreticSecurityProof2005},
conditioned on acceptance:
\begin{align}
    \epsilon_{\mathrm{cor}}
    &:=\Pr[K_A\ne K_B\mid\mathrm{pass}],
      \label{eq:correctness}\\
    \epsilon_{\mathrm{sec}}
    &:=\frac12\left\|
      \sigma_{K_AE^nT\mid\mathrm{pass}}
      -\tau_\ell\otimes\sigma_{E^nT\mid\mathrm{pass}}
      \right\|_1. \label{eq:secrecy}
\end{align}
The uniform key state is
$\tau_\ell=2^{-\ell}\sum_{k\in\{0,1\}^{\ell}}\lvert k\rangle\langle k\rvert$.
For an operator $M$, the trace norm is
$\|M\|_1=\Tr\sqrt{M^\dagger M}$; half the trace norm of a difference of
states is their trace distance.

\begin{theorem}[Classical postprocessing converse]
\label{thm:bound_secrecy}
Let $q\in[q^*,1/2]$, where $q^*=(5-\sqrt5)/10$. Every protocol described
above, acting on $\omega_{XYE}^{\otimes n}$, satisfies
\begin{equation}
    \boxed{\epsilon_{\mathrm{sec}}+4\epsilon_{\mathrm{cor}}
    \ge\frac12}. \label{eq:finite_block_bound}
\end{equation}
Consequently, the correctness and conditional secrecy errors cannot both tend to zero, and the measured source has zero secret-key capacity under classical public postprocessing.
\end{theorem}

\begin{proof}
The matrix entries of \cref{eq:quantum_source} imply that
$\operatorname{span}\{\lvert e_{00}\rangle,\lvert e_{11}\rangle\}$ is orthogonal to
$\operatorname{span}\{\lvert e_{01}\rangle,\lvert e_{10}\rangle\}$. The overlap
of the equal-bit states is
\begin{equation}
    s:=\langle e_{00}\mid e_{11}\rangle
       =\frac{1-2q}{1-q}. \label{eq:overlap}
\end{equation}
For the assumed range of $q$, we have
$0\le s\le(\sqrt5-1)/2<1$,
and hence $s\le 1-s^2$.

\emph{1. Auxiliary conditional independence.}
Following the auxiliary-variable comparison
approach~\cite{gohariCodingPositiveRate2020,
pauwelsBipartiteBoundInformation2026}, introduce a fictitious
random variable $J$ with alphabet $\mathcal J:=\{0,1,\bot\}$,
where $\bot$ is an auxiliary flag. Define its joint distribution with $X,Y$ as follows.
With probability $1-2q$, choose a uniform bit $J$ and set $X=Y=J$.
With probability $2q$, set $J=\bot$ and choose $X,Y$ independently and
uniformly. This reproduces the marginal probabilities in
\cref{eq:ccq_state}. Given $J=0$ or $1$, both honest bits are fixed;
given $J=\bot$, they are independent. Thus
\begin{equation}
    X\indep Y\mid J, \label{eq:aux_independence}
\end{equation}
where $\indep$ denotes conditional statistical independence.

Write $Z:=(X,Y)$ for the raw pair, $z=(x,y)$ for one of its values, and
$\lvert e_z\rangle:=\lvert e_{xy}\rangle$. The auxiliary conditional law
$W(j\mid z):=\Pr[J=j\mid Z=z]$, for $j\in\mathcal J$, is
\begin{align}
    W(b\mid(b,b))&=s,
    &W(\bot\mid(b,b))&=1-s, \nonumber\\
    W(\bot\mid(x,y))&=1\quad(x\ne y),
    \label{eq:aux_channel}
\end{align}
with all unspecified probabilities zero.

\emph{2. Transfer of estimators to Eve.}
The moment comparison below is related to methods for bounding
quantum private capacity through observable reconstruction and
binary estimation~\cite{zhuQuantumIncapacityObservable2026,
hircheQuantumEstimationChannel2026}.
Define the reciprocal vectors and Hermitian operators
\begin{align}
    \lvert v_b\rangle
    &:=\frac{\lvert e_{bb}\rangle-s\lvert e_{1-b,1-b}\rangle}{1-s^2},
       \qquad b\in\{0,1\}, \nonumber\\
    F_b&:=s\lvert v_b\rangle\langle v_b\rvert,
    \qquad F_\bot:=I_E-F_0-F_1, \label{eq:lift_operators}
\end{align}
where $I_E$ is the identity on $\mathcal H_E$. They satisfy
\begin{align}
    \langle v_b\mid e_{bb}\rangle&=1,
    &\langle v_b\mid e_{1-b,1-b}\rangle&=0, \nonumber\\
    \|v_b\|^2&=\frac1{1-s^2},
    &\langle e_z\lvert F_j\rvert e_z\rangle&=W(j\mid z),
    \label{eq:lift_means}
\end{align}
where $\|\cdot\|$ denotes the Hilbert-space norm.

For a function $f:\mathcal J\to\mathbb C$, define
\begin{align}
    \mathcal C_z f&:=\sum_{j\in\mathcal J}f(j)F_j\lvert e_z\rangle,
       \nonumber\\
    \|f\|_z^2&:=\sum_{j\in\mathcal J}W(j\mid z)|f(j)|^2.
       \label{eq:weighted_norm}
\end{align}
The latter is the squared norm in the weighted space
$L^2(W(\cdot\mid z))$, in which functions agreeing wherever
$W(j\mid z)>0$ are identified. For an equal pair $z=(b,b)$, direct
calculation gives
\begin{align}
    \|\mathcal C_{(b,b)}f\|^2
    &=s|f(b)|^2+(1-s)|f(\bot)|^2 \nonumber\\
    &\quad-s\left(1-\frac{s}{1-s^2}\right)
       |f(b)-f(\bot)|^2 \nonumber\\
    &\le\|f\|_{(b,b)}^2. \label{eq:scalar_contraction}
\end{align}
For an unequal pair, $\mathcal C_zf=f(\bot)\lvert e_z\rangle$, so
$\|\mathcal C_zf\|^2=\|f\|_z^2$. Hence every $\mathcal C_z$ is a
contraction from $L^2(W(\cdot\mid z))$ to $\mathcal H_E$. These inequalities
also show that zero-norm functions map to zero, so the maps are well defined.

Now let $Z^n=(Z_1,\ldots,Z_n)$, where $Z_i=(X_i,Y_i)$, and write
$z^n=(z_1,\ldots,z_n)$ for its values. Conditional on $Z^n=z^n$, give
$J^n=(J_1,\ldots,J_n)$ the product law
\begin{equation}
    W^{\otimes n}(j^n\mid z^n):=\prod_{i=1}^n W(j_i\mid z_i),
    \label{eq:aux_product}
\end{equation}
where $j^n=(j_1,\ldots,j_n)\in\mathcal J^n$. For every real function
$g:\mathcal J^n\to\mathbb R$, set
\begin{align}
    H_g&:=\sum_{j^n\in\mathcal J^n}g(j^n)
       \bigotimes_{i=1}^n F_{j_i}, \nonumber\\
    \lvert e_{z^n}\rangle&:=\bigotimes_{i=1}^n\lvert e_{z_i}\rangle.
       \label{eq:block_observable}
\end{align}
The operator $H_g$ is Hermitian on $\mathcal H_E^{\otimes n}$.
The product weighted function space is the tensor product of the one-copy
spaces. Since tensor products of contractions are contractions,
\begin{align}
    \langle e_{z^n}\lvert H_g\rvert e_{z^n}\rangle
    &=\mathbb E[g(J^n)\mid Z^n=z^n],
       \label{eq:block_first_moment}\\
    \langle e_{z^n}\lvert H_g^2\rvert e_{z^n}\rangle
    &=\left\|\left(\bigotimes_{i=1}^n\mathcal C_{z_i}\right)g\right\|^2
       \nonumber\\
    &\le\mathbb E[g(J^n)^2\mid Z^n=z^n].
       \label{eq:block_second_moment}
\end{align}
Here $\mathbb E$ denotes expectation with respect to the stated conditional
law. The operators $F_j$ need not be positive and are not assumed to form a
measurement; Eve will measure the Hermitian observable $H_g$.

\emph{3. Application to the final key.}
To apply this comparison to a protocol, sample $J^n$ from
\cref{eq:aux_product} using randomness independent of all protocol
randomness. Let $R_A,R_B$ collect Alice's and Bob's private random choices.
Their initial local views
\begin{equation}
    U:=(X^n,R_A),\qquad V:=(Y^n,R_B)
\end{equation}
then satisfy $U\indep V\mid J^n$. Fixing the independent public randomness
preserves this property. With the previous transcript fixed, each new
message is a function of the speaker's initial local view, so conditioning
on that message restricts only that view. Conditional independence therefore survives every message, including
public stopping and acceptance decisions. Thus
$U\perp\!\!\!\perp V\mid(J^n,T)$, by the rectangle property
of classical communication protocols~\cite[Lemma~6.7]{barYossef2004Information}.

Let $K_{A,1}$ and $K_{B,1}$ be the first bits of the output keys and define
their signs by
\begin{equation}
    \kappa_A:=(-1)^{K_{A,1}},\qquad
    \kappa_B:=(-1)^{K_{B,1}}.
\end{equation}
Both signs lie in $\{-1,+1\}$. All remaining probabilities and expectations
are conditioned on $\mathrm{pass}$. Since each output depends only on its
owner's view and the transcript,
$\kappa_A\indep\kappa_B\mid(J^n,T)$. For an accepted transcript value $t$,
define
\begin{equation}
    g_t(j^n):=\mathbb E[\kappa_B\mid J^n=j^n,T=t],
    \label{eq:key_estimator}
\end{equation}
and assign any value in $[-1,1]$ on zero-probability branches. Then
$|g_t(j^n)|\le1$, and conditional independence gives
\begin{equation}
    \mathbb E[\kappa_A g_T(J^n)]
    =\mathbb E[\kappa_A\kappa_B],
    \label{eq:key_correlation}
\end{equation}
where $g_T$ denotes $g_t$ evaluated at the observed transcript.

After learning $T=t$, Eve measures $H_t:=H_{g_t}$ spectrally and obtains
a real outcome $h$. Conditional on the entire original raw string
$Z^n=z^n$, further conditioning on the private tapes, keys and transcript
changes neither Eve's state $\lvert e_{z^n}\rangle\langle e_{z^n}\rvert$
nor the auxiliary law \cref{eq:aux_product}. We may therefore average
\cref{eq:block_first_moment,eq:block_second_moment} over accepted runs.
Using \cref{eq:key_correlation} and $|g_t|\le1$ gives
\begin{align}
    \mathbb E[(\kappa_A-h)^2]
    &=1-2\mathbb E[\kappa_Ah]+\mathbb E[h^2] \nonumber\\
    &\le1-2\mathbb E[\kappa_Ag_T(J^n)]
       +\mathbb E[g_T(J^n)^2] \nonumber\\
    &\le2-2\mathbb E[\kappa_A\kappa_B] \nonumber\\
    &=4\Pr[\kappa_A\ne\kappa_B]
      \le4\epsilon_{\mathrm{cor}}.
      \label{eq:key_mse}
\end{align}
The last inequality holds because first-bit disagreement implies full-key
disagreement.

Eve's guess $\widehat\kappa_E$ is $+1$ when $h\ge0$ and $-1$ otherwise.
A wrong guess implies $(\kappa_A-h)^2\ge1$, so
\begin{equation}
    \Pr[\widehat\kappa_E=\kappa_A]\ge1-4\epsilon_{\mathrm{cor}}.
\end{equation}
For the ideal accepted state in \cref{eq:secrecy}, the first key bit is
uniform and independent of Eve's system and $T$, so the same guessing
procedure succeeds with probability exactly $1/2$. Trace distance bounds
the difference between these success probabilities, including after
discarding the other key bits. Therefore
\begin{equation}
    \epsilon_{\mathrm{sec}}
    \ge\Pr[\widehat\kappa_E=\kappa_A]-\frac12
    \ge\frac12-4\epsilon_{\mathrm{cor}},
\end{equation}
which proves \cref{thm:bound_secrecy}.
\end{proof}

If the output length is public and variable, apply the same argument to the
first output bit conditional on accepting with a nonempty key.

\textit{Conclusion.---}
The compatible collective attack above prevents reliable and conditionally
secret key generation for $q\ge q^*$ under arbitrary classical
postprocessing of the measured source. Together with known achievability
below $q^*$, this fixes the classical postprocessing threshold. In the
interval $q^*\le q<1/3$, the underlying two-qubit state remains entangled,
while its measured data exhibit bound secrecy.

\textit{Acknowledgements.---}
We acknowledge partial support from Japan's Council for Science,
Technology and Innovation (CSTI) through the Cross-ministerial Strategic
Innovation Promotion Program (SIP) project ``Promoting the application of
advanced quantum technology platforms to social issues'' (QST grant
JPJ012367).

The main argument in the proof of \cref{thm:bound_secrecy} was obtained with the help of ChatGPT 6.1 (OpenAI). We have checked and rewritten all arguments in the manuscript and take full responsibility for their validity.

\bibliography{ref}

\end{document}